# An ERP Implementation Method : Studying a Pharmaceutical Company

Emmanouil Kolezakis

The Greek Ministry of Education

ekolezakis@sch.gr

Abstract: Analysing the development process for an ERP solution, in our case SAP, is one of the most critical processes in implementing standard software packages. Modelling of the proposed system can facilitate the development of enterprise systems not from scratch but through use of predefined parts who represents the best knowledge captured from numerous case studies. This aim at abstracting the specification of the required information system as well as modelling the process towards this goal. Modelling plays a central role in the organisation of the information systems development process and the information systems community has developed a large number of conceptual models, systems of concepts, for representing conceptual schemata. In the area of ERP systems, because of the characteristics that distinguishes them, conceptual modelling can help in all aspects of the development process, from goal elicitation to reuse of the captured knowledge, through the use of the appropriate modelling schemata. SAP offers a standardised software solution, thus making easier the alignment of SAP requirements to enterprise requirements in a goal form, and the correspondent business processes.

## 1. Introduction

ERP systems provide a software solution comprising several interconnected modules covering most of the key functions. For example SAP has modules for human-resource, material logistics, treasury, etc.This paper proposes a framework for the treatment of goal acquisition, alignment and reuse within the enterprise in order to facilitate the use of SAP. In the following section the ontology of the Reusable Organisational Change (ROC) framework is presented through the case of Electro Tech.

 "Electro Tech is a fictional company, created from the collection of a variety of true life business practices" [Hiquet 1998]. It is manufacturer of electrical components and factory automation products. It has been in the business since late 50s and has demonstrated a consistent growth during 60s, 70s and 80s. The problems it faces are not unique but typical of a company who come across, IT evolution, globalisation, integration, mergers etc.

## 2. Ontology of the Methodology

The ROC framework consists of four static affinities namely:
(1) Organisational Goals
(2) Business Process Models
(3) Project Deliverables
(4) Requirement Reuse Plan.

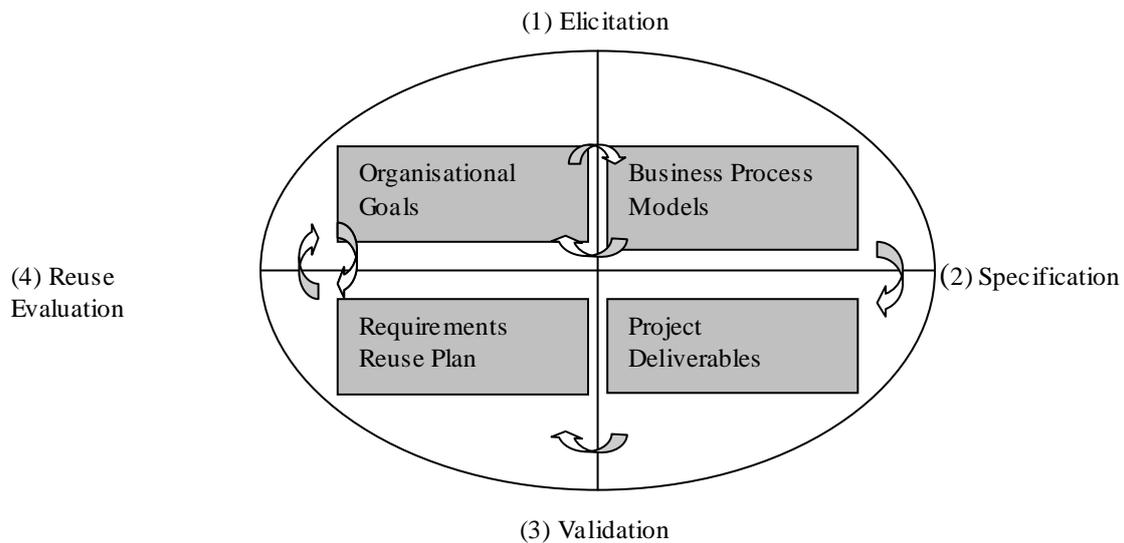

Figure 1: ROC Application Process

It consists of four dynamic affinities as well:
(1) Elicitation
(2) Specification
(3) Validation
(4) Reuse Evaluation

The elicitation phase of the framework includes the determination of the high level objectives of the enterprise as well as the more functional one. Product of this phase is the goal-graph notation (see fig.3).

In the second stage the outcome is the Petri-net notation of the As-Is state. In the third stage we can see the alignment of the processes of both SAP and enterprise. In this stage we have a more concrete idea of what to expect from the project itself. Product of this stage is the Petri-net notation (The To-Be state).

Finally the Reuse Evaluation phase stores the knowledge captured during the project implementation process for future use. Especially stores the strategy followed during the third phase of the framework so that we use this knowledge in possible similar forthcoming projects.

## 2.1 Goal Elicitation Sub-model

The goal-elicitation sub-model is illustrated in fig.2. Central to this view is the concept of organisational goal. An organisational goal is a desired state of affairs that needs to be attained. Typical goals in the case of Electro Tech are: "*improve IS services*" or "*Automate payroll*" or "*satisfy customer need for information from their suppliers*" etc.

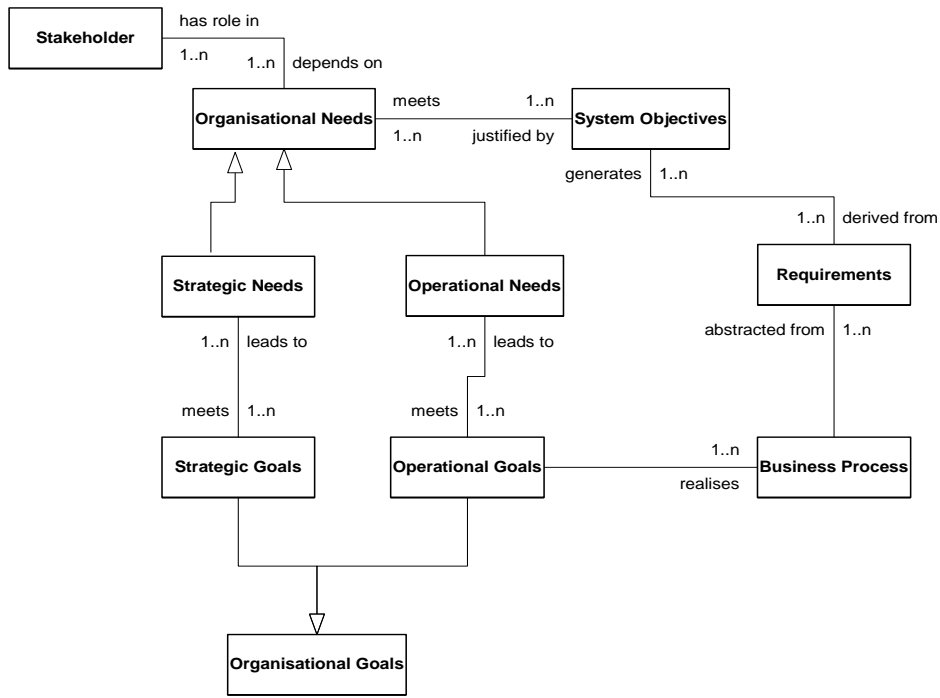

Figure 2: Goal Elicitation Sub-model

Goals are pertain to stakeholders. A stakeholder is defined as someone who has an interest in the system design and usage. Examples of stakeholders are: managers, system designers, system users, customers etc.

Systems are built to primarily satisfy organisational needs. These needs may either be long term strategic needs or more immediate operational needs. These two needs support each other and are often more detailed in a goal hierarchy. We must distinguish between goals and needs. Goals derived from needs which are expressed in a more abstract manner. For example "*need for information*" is a general organisational need for Electro Tech. From this need derived the goal "*automate payroll*" and "*satisfy customer need for information from suppliers*"

System objectives elicited from organisational needs and are determined by stakeholders. For example the objective: "*supply with the latest technology*" is an objective of the system determined by stakeholders, probably management but depends on organisational needs for leading edge technology. Operational needs lead to the formulation of operational goals and strategic needs to the formulation of strategic goals. For example "*Buy a VP of sales and marketing state of the art system*" is a strategic goal whereas "*Buy a VP of sales and marketing system using lotus 1-2-3*" is an operational one.

The requirements for the system are generated from system objectives because high level goals are too vague to be called requirements. Requirements must be more specific to proceed further.

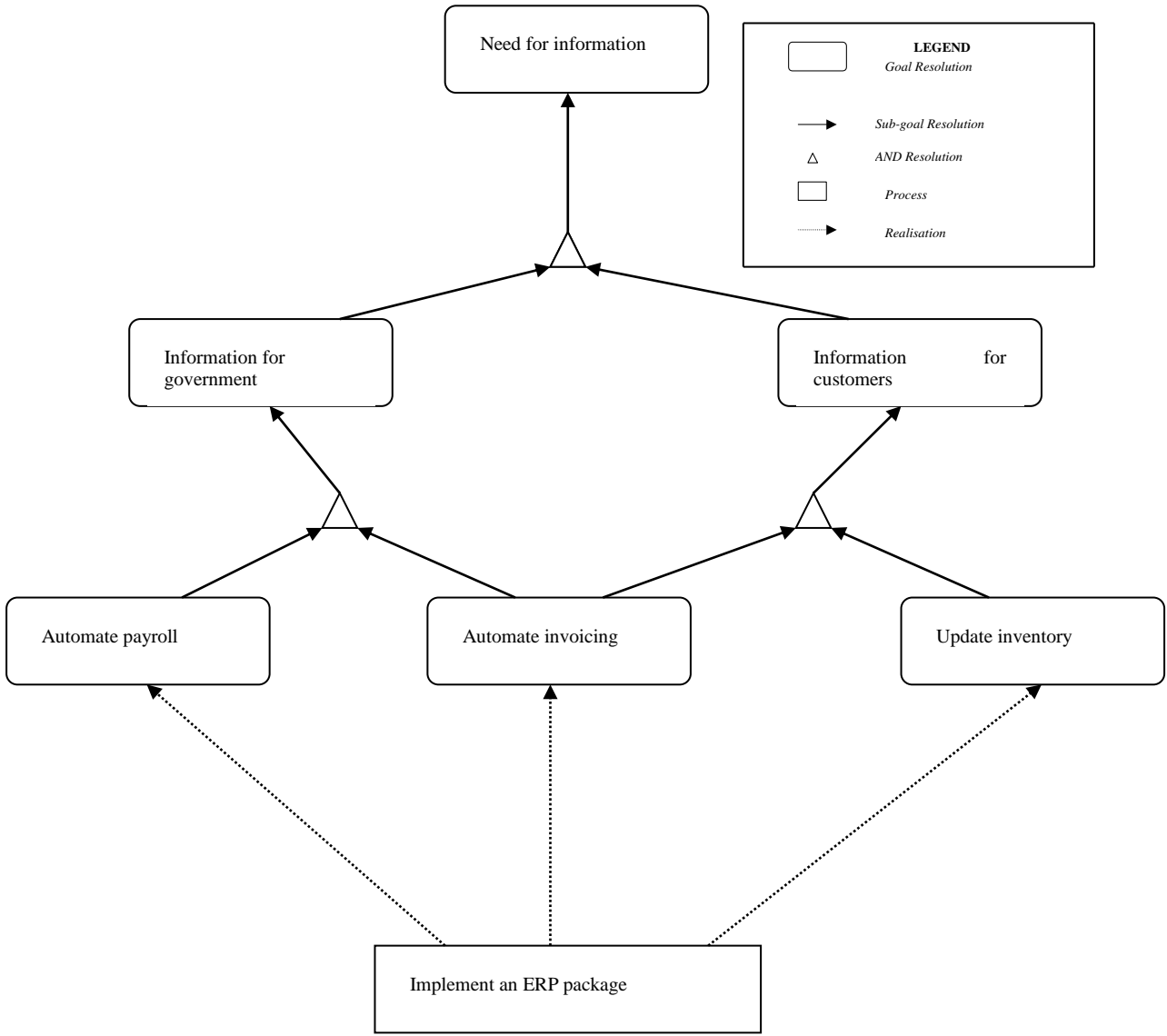

Figure 3: Goal-Graph for Electro-Tech

In fig. 3 we have an example of the elicitation of the enterprise goals in a notation as it has been used from [Loucopoulos et Kavakli 1997]. We start from the high level objectives of the enterprise such as "*need for information*" and we end with the low level objectives, more concrete goals such as "*autmate payroll*", "*automate invoicing*" and "*update inventory*". These goals could lead as it has been stated before into operational goals such as "*Buy a VP of sales and marketing system using lotus 1-2-3*". This goal graph depicts mainly the strategic goals derived from strategic needs. The realisation of these goals comes from the process "implement an ERP package" in our case SAP. If we proceed further from the strategic goals we have the elicitation of the operational goals. The operational goals are realised by the business.

## 2.2 Specification of the Business Processes Sub-model

The specification sub-model is illustrated in fig.4. Central to this meta-model are the change goals. Change goals are elicited from current organisational goals and with the help of Business Process Models.

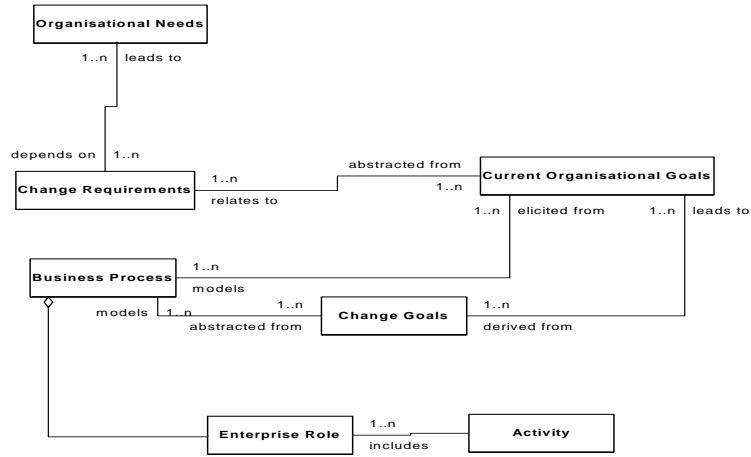

Figure 4: Specification Sub-model.

.

Organisational needs leads to change requirements. For example the "*need for an integrated IS because of very diverse citation*" leads to the change requirement "*improve MIS services*". That means that we model the business process then the goals are derived from the models and are specified in Petri-nets notation. A current organisational goal can be "*satisfy customer need for information from suppliers*" and a change goal "*develop a homegrown IS*".

## 2.3 Validation Sub-model

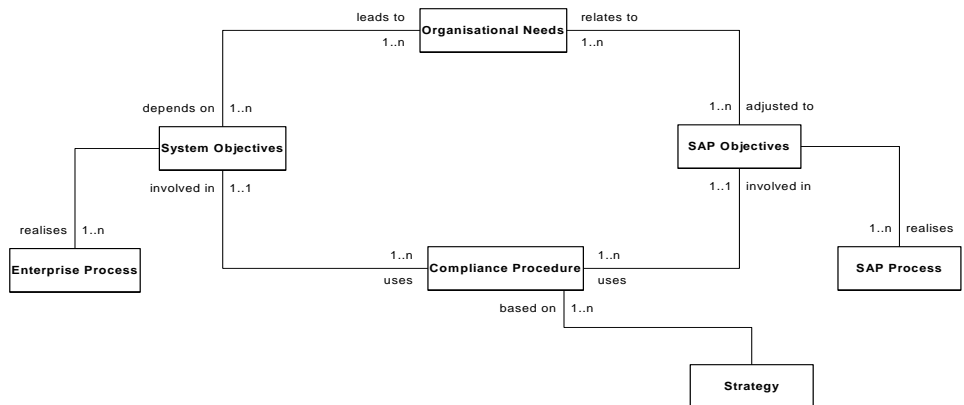

Figure 5: Validation Sub-model

In this stage it is being examined if the business objectives can be fulfilled and what changes must be made in the business processes of the enterprise in order to realise these. This is an interactive process that transforms the enterprise's business requirements into a future SAP solution. This interactive process provides continuous feedback mechanisms that initially identify gaps and then evolve to filling the gaps.

Installations of ERP systems are difficult to align to specific requirements of the enterprise because of the low level at which functionality is described. Central to this metamodel is the compliance procedure. The following terms are defined:
SAP goals: The tasks carried out by a SAP function.
SAP process: The process who realises a SAP goal.
SAP strategy: The combination of all necessary SAP processes, in order to reach a SAP goal.

The main reason for thinking in terms of goals (intentional level) and strategies (Strategy level) is that we need a common way of communication between SAP and enterprise. Organisations think in terms of their objectives and their strategies and SAP functions have a supportive role. SAP goals must support and implement enterprise goals.

## 2.4 Reuse Evaluation

The purpose is to reduce significantly the task of creating application-specific models and systems: the user select relevant parts of the reference model, adapts them to the problem at hand, and configures an overall solution from these adapted parts. Since the analysis of a domain can take an enormous effort when started from scratch, the use of reference models has been reported to save up to 80% in development costs for systems in standardised domains [Scheer 1994]. According to Ramesh and Jarke [Ramesh and Jarke 2001] reference models have become highly successful in many industries and among the best known examples is the SAP approach. Each case can be consider a scenario S. The retrieval of a scenario S can follow the algorithm [Aamodt and Plaza 1994]:

A. **Problem create New Case**
B. **Retrieve an old similar Case**
C. **Compare Retrieved Case and New Case**
D. **Reuse Solved Case**
E. **Test Suggested Solution**
F. **Retain Learned Case**
 Next

## 3. Aligning ERP to Enterprise

In this section we examine and model in more detail the stage between (1) elicitation and (3) validation of the framework whereas the alignment of both SAP and enterprise processes takes place and the most vital work is being done. We study the current state of the enterprise (As-Is) and then we study the desired state (To-Be) of the proposed SAP solution. Both states are representing in Petri-nets, a modelling element where the correspondent strategy for each transition is the key feature for the alignment.

## 3.1 Logical Organisation

The levels of abstraction are (Figure 6):

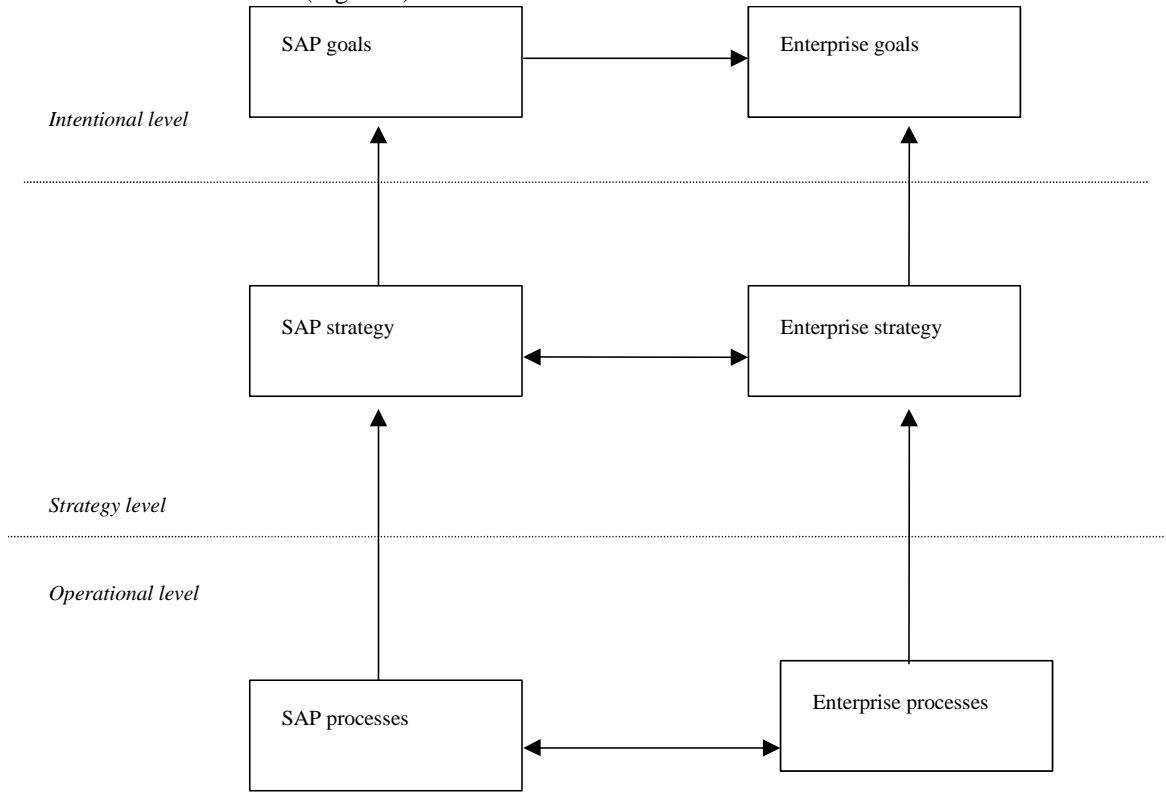

Figure 6: Levels of Abstraction

(a) Intentional level
(b) Strategy level
(c) Operational level

The Intentional level defines concepts such as goals. The SAP goals must be supportive to the enterprise goals and their realization must help towards the realization of the enterprise objectives.

The Strategy level defines concepts such as SAP strategy or enterprise strategy.

The Operational level includes concepts such as SAP processes or enterprise process, namely the process that realizes the enterprise goals.

So we have alignment of the processes in a strategy level, expressed in Petri-nets and the SAP goals support enterprise's high level objectives.

## 3.2 Modelling the Current State

Petri-nets [Petri 1962] are to model the current state of the enterprise, in our case Electro Tech. The production planning module (PP) of SAP is used as an example. The PP module is a flexible module containing several alternative strategies adjusted to each particular enterprise according to its objective and the targeted process the stakeholders want to implement. Because of the modularity which distinguishes SAP the Petri-net notation is a particularly suitable modeling tool for ERP systems in our case SAP.

Each node in the Petri-net notation corresponds to a state in the production planning process. Using the triplet form <source state, target state, strategy> we can have the correspondent textual notation for the transmission from a place to another place. This graphical notation used to represent the correspondent fragments of the SAP production planning process, having same milestones but probably different strategy can be used for adjusting the SAP planning strategy into the targeted enterprise process. The current status of the enterprise as it is described is [Hiquet 1998]:

(a) There is a serious problem in the control of approve vendors which is fragmented and controlled manually. The manual purchasing system can cause errors at the first place and at the second, there is no history option or any possibility for anticipation. As a result the Sales and Operations plan is created manually.
(b) There is no demand management strategy in the supply of raw material. That means no forecasting strategy within the supply chain.
(c) There is no real time production planning strategy. That means need for production in advance and application of a stock strategy. (Make-to-Stock strategy).
(d) Lack of an on-line order processing system even automated.

As it comes up from the previous conclusions the identified problems are fall into two categories:

- Supply Chain Management problems, for example (a) and (b),
- Demand Side problems, for example © and (d).

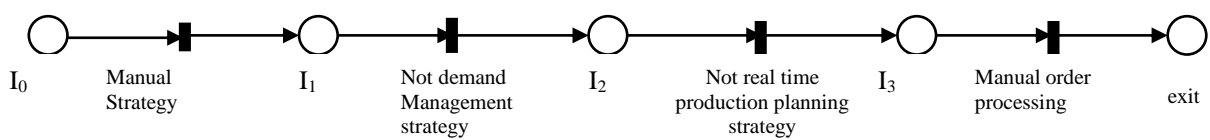

Figure 7: The Current Status of Electro Tech.

$I_0$: start, $I_1$: support material, $I_2$: work with material, $I_3$: Stock.

Using the notation described previously we can depict all the above making thus possible the comparison or better the adjustment of the future SAP solution into business processes of the enterprise (fig.7)

Each fragment of the above representation corresponds to a business situation, illustrating the problems stated. We use the triplet form <source state, target state, strategy> to represent the transition from one place to another place, and the strategy element depicts how this is being done.

In the current status of Electro Tech the fragments are:

**PF$_1$** :<(start), (support material), *manual strategy*>,
**PF$_2$** :<(support material), (work with material), *Not demand management strategy*>,
**PF$_3$** :<(work with material), (stock), *Not real time production planning strategy*>,
**PF$_4$** :<(Stock), exit, *manual order processing strategy*>.

In a few words we can describe the functionality of the company as:

**PF₁** :<(start), (support material), *manual strategy*>: Production scheduling is reacting to the occurrences in the plant. All planning strategy is based on word-of-mouth.

**PF₂** :< (support material), (work with material), *not demand management strategy*>:
As commonly defined, demand management is the function of recognising all demands for a product, including forecasts, customer orders, interplant orders, etc. The output of demand management is referred to as the demand program. It consists of a list of independent requirements for each material specifying the quantities and dates the material is needed. Multiple versions of the planned independent requirements can be useful for segregating and managing different components of demand. In Electro Tech the sales demand is composed of direct sales and warranty replacement. The warranty replacement demand is calculated manually based on reliability data and total number of units in service.

**PF₃** :<(work with material ), (stock), *not real time production planning strategy*>: As a result of the previous problems there are delays in the production process of Electro Tech. Especially when a company has several plants depended one to another as far as the production process is concerned.

**PF₄** :<(stock), exit, *manual order processing strategy*>: What is really needed is a truly integrated system, without redundancy and with real-time transactional integration between business functions.

The correspondence between the identified problems and the process fragments is:

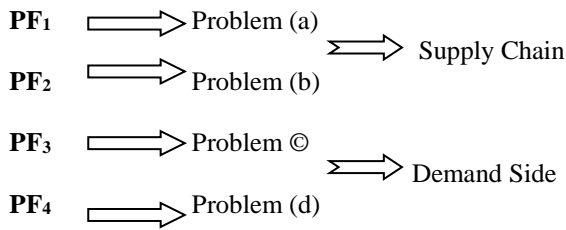

**PF₁** ⇒ Problem (a)
⇒ Supply Chain
**PF₂** ⇒ Problem (b)

**PF₃** ⇒ Problem ©
⇒ Demand Side
**PF₄** ⇒ Problem (d)

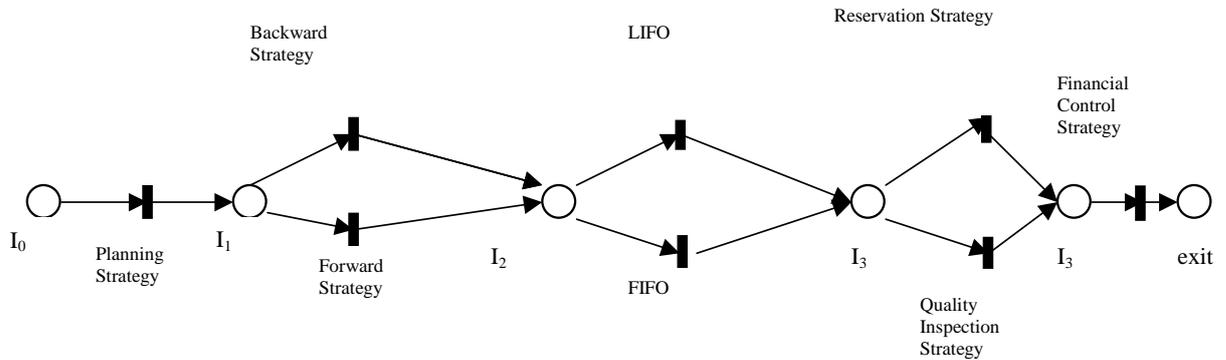

Figure 8: Proposed SAP Solution

## 3.3 Modelling the Proposed System

In the same manner we modeled the current state of the enterprise we model the future SAP solution. Because we are referred to Production Planning process and the Production Planning module of SAP is a highly flexible module consisting of several implementation strategies we must select an appropriate implementation strategy for the particular company. It is difficult to present a general overview of the SAP production planning process. The production planning functionality is a highly flexible collection of sub-modules and functionality that can be linked together to form a coherent planning and scheduling process. Every SAP implementation has the opportunity to use all or a part of the SAP planning functionality in order to meet organisations specific planning needs [Keller and Teufel 1998].

Figure 8 depicts an example of how the functional sub-modules or according to our determination components can be used in order the specific planning needs of the company to be fulfilled. In this example the sales representative create sales plan in the sales and operations component. These sales plans are then copied into Demand Management by the master production scheduler and smoothed out from weekly basis into daily basis. Once the master plan (in Demand Management) is satisfactory, master production scheduling is performed, followed by detailed material requirements planning.

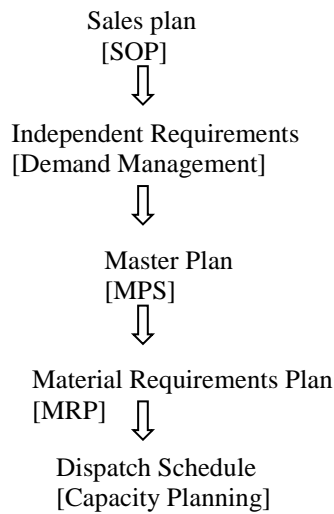

Figure 9: Sample SAP production planning process [Hiquet 1998].

The detailed material requirements are fed into the capacity planning system for finite production scheduling and dispatching.

This process would be applicable for a company that has accurate sales forecast information based on sales account representative feedback from customers, or some form of accurate market predictors. The production environment would typically be a job shop with relatively expensive finished goods being produced in a complex manufacturing and assembly process.

For Electro Tech we selected a make-to-stock strategy consisting of the Sales and Operations component, Master Processing Scheduling, Material Requirements Planning (MRP), Quality Management (QM), and Product Costing (PC) component. The representation in Petri-nets of the above module is the one in figure 8 whereas $I_0$:start, $I_1$:support material, $I_2$:work with material, $I_3$:Stock.

The business process in SAP concerning the PP module with the above planning strategy has the following fragments:

**PF$_1$**<(start), (Support material), *planning strategy*>
**PF$_2$**<(Support material), (work with material), *backward strategy*>
**PF$_3$**<(Support material), (Work with material), *forward strategy*>
**PF$_4$**<(Work with material), (stock product), *LIFO*>
**PF$_5$**<(Work with material), (stock product), *FIFO*>
**PF$_6$**<(Stock Product), (Stock Product), *Reservation Strategy*>
**PF$_7$**<(Stock Product), (Stock Product), *Quality Inspection Strategy*>
**PF$_8$**<(Stock Product), exit, *Financial Control Strategy*>.

Each of the above fragments treats a specific problem of the enterprise. For example the fragments PF1, PF2, PF3, PF4, PF5 resolve problems related to supply chain, whereas the fragments PF6, PF7, PF8 resolve problems related to demand side of the enterprise. The adjustment of the components of the Production Planning module of SAP has been done in such a manner so that to implement the selected planning strategy. Diagrammatically this can be proved as it is depicted below.

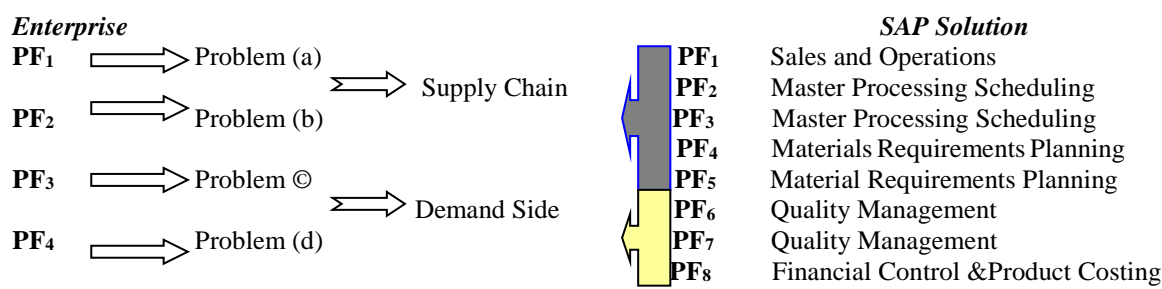

## 4. Study

We have seen so far the methodology and the ontology of the ROC framework for assisting mainly the alignment of the enterprise requirements to SAP requirements and the reuse of the captured knowledge. This section presents the empirical results and observations from applying the above framework in an industrial application capture from the literature.

The intention of this discussion is twofold: The first is to assess the applicability of the ROC framework on a non-trivial application. The second is to discuss a number of challenging issues that need to be addressed in order the proposed Framework to support real, complex tasks in the context of business modelling and reuse within the ERP domain and particularly SAP.

### 4.1 The Pharmaceutical Industry

The application considered is that of the implementation of SAP R/3 in a large pharmaceutical company. "*This case highlights initial mistakes during this journey, strategies that helped overcome these mistakes, and how R/3 delivered operational efficiencies and competitive advantage under difficult business circumstances.*" [Bhattacherjee 2000].

We concentrate in the production planning module of SAP and thus we study the reengineering of the manufacturing process of the company. The considered case study was selected because it represents a unique real industrial application. The company is a typical example of today's multinational companies that went through acquisitions and mergers to reach today's status.

Following the application process of the ROC framework, firstly we consider the goals of the company as they stated from the stakeholders. A goal graph is created depicted the targets of the company and their abstraction from intentional features to operational ones.

### 4.1.1 Application Background.

Geneva pharmaceuticals, Inc. is one of the largest generic drug manufacturers and the North American hub for the generic drugs division of Swiss pharmaceutical and life sciences company Novartis International AG. Geneva's primary business processes are manufacturing and distribution. Its manufacturing process is scientific, controlled, and highly precise. Until 1996, Geneva's information systems consisted of multiple software programs for managing mission-critical functions such as procurement, manufacturing planning, accounting, and sales. The systems infrastructure was predominantly oriented around IBM's technologies. The primary hardware platform was a mid-range IBM AS/400. The software platform was based primarily on IBM's DB/2 database. Desktop microcomputers were connected to the AS/400 via a token-ring local area network.

Business units funded and deployed applications as needed in an ad hoc manner, without concern for maintenance or enterprise-wide interoperability. Data shared across business units (e.g., accounts receivable data was used by both order management and financial accounting, customer demand was used both in sales and manufacturing) were double booked and re-keyed manually. The above configuration led to data entry errors, error processing costs and data inconsistency. Furthermore the above status did not support value-added processes that cut across the enterprise, for example end-to-end supply chain management. It was apparent that a common, integrated company-wide solution was needed to improve data accuracy and consistency, reduce maintenance costs and enable value-added processes. The acquisition of the enterprise goals is illustrated in fig. 10. The main objective of the company is twofold: growth via acquisition thus balancing manufacturing and purchasing, and internal growth through the reduce of operational cost.

out by SAP functions are supportive to high-level enterprise goals or more specific are sub-goals of the high-level enterprise goals. The alignment of the SAP processes to enterprise processes is being done according to ROC framework in a strategy level carried out in Petri-nets.

The alignment in this case is being done from the SAP side towards the goals of the enterprise. For example we don't have the implementation of SAP modules as a unity but the adjustment of different SAP components to targeted business processes of the enterprise (fig. 11).

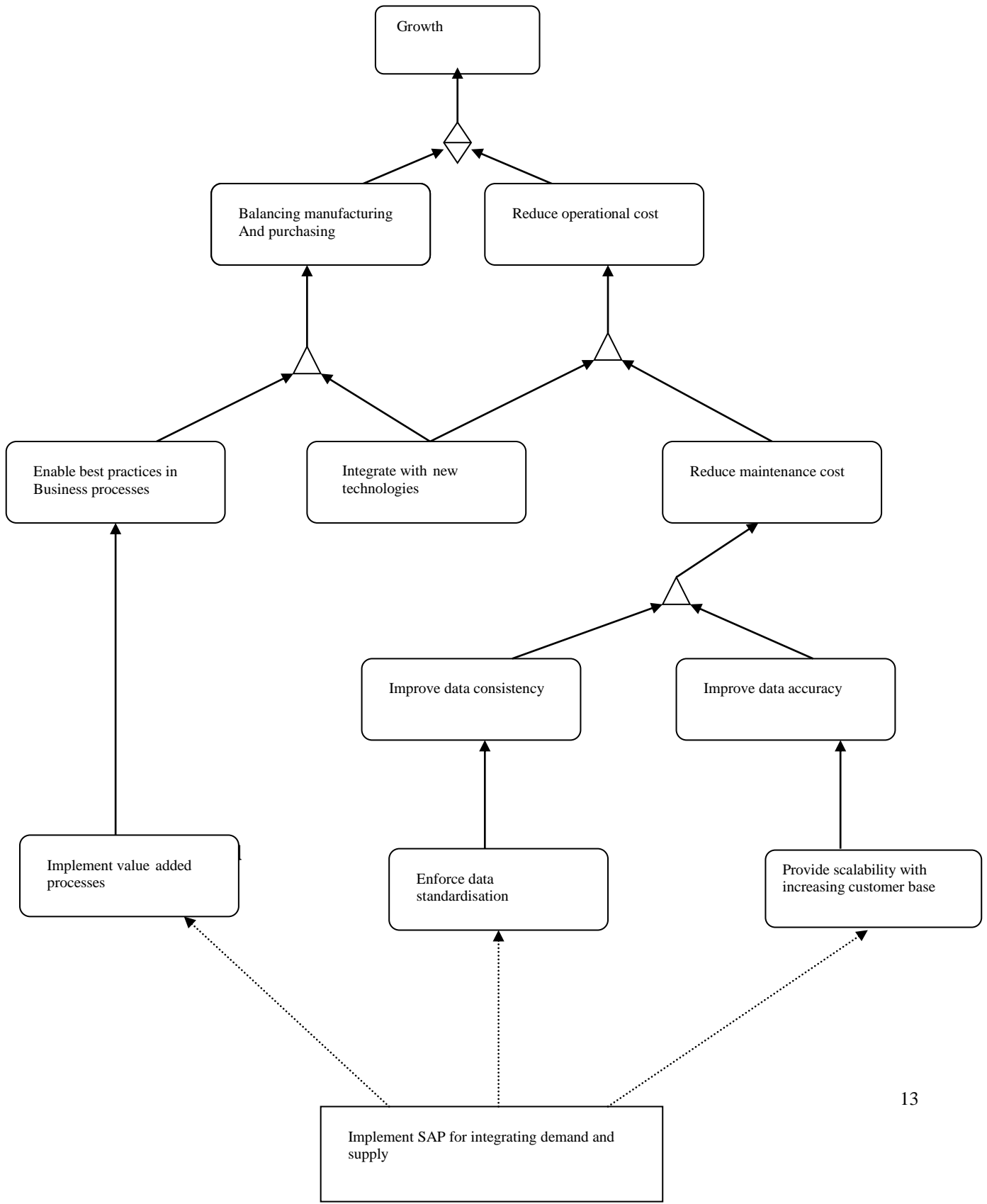

Figure 10: Abstraction of Geneva's goals

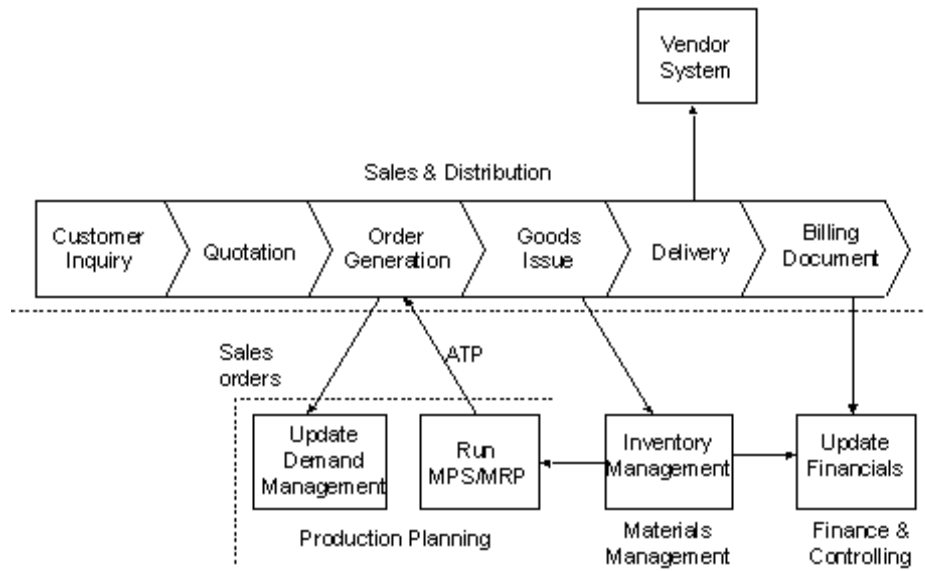

Figure 11: Targeted Business process for order management [Battacherjee 2000]

## 4.1.2 Demand side implementation

The main target of demand side implementation was to redesign demand side processes such as marketing and sales, order fulfillment, customer service, accounts receivable, and implementing the reengineered processes using SAP. As it is depicted from Fig. 2 two components of PP module are implemented as far as the order management process of the enterprise is concerned. These are demand management and MPS/MRP (Master Production Scheduling/Materials Requirements Planning). Moreover they are implemented inventory management from MM (Material Management) module and Financing and control. The current situation of the enterprise will be as follows:

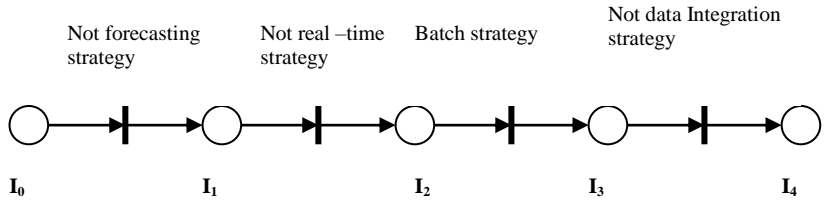

Figure 12: Enterprise order management process (AS-IS)

The states are the following: $I_0$: **Customer inquiry,** $I_1$: **Order generation,** $I_2$: **Goods issue,** $I_3$: **Goods delivery,** $I_4$: **Billing.**

The main problems identified as they are extracted from the representation of the current status of the enterprise are: Before the initiation of the SAP project there were not forecasting strategy for all the customers. The only existed forecasting strategy was concerning key customers accounts. In the whole process there was a shortage of data integration and real time access capabilities. Prior to SAP orders were coming once a night, chargebacks were coming once a day, invoicing was done overnight, shipments were getting posted once a day.

The SAP process for order management is the following:

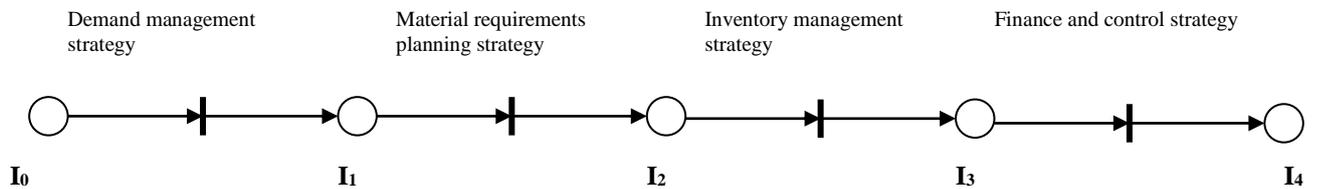

Figure 13: SAP order management process

Whereas $I_0$: **Customer inquiry,** $I_1$: **Order generation,** $I_2$: **Goods issue,** $I_3$: **Goods delivery,** $I_4$: **Billing.**

The alignment of these two processes for order management generates the targeted process of fig.11. This representation in Petri-nets helps us to distinguish which components should be implemented. This representation is a high-level one but we could go in a lower level through refinement.

The alignment will be done in the triplet form <(source state), (target state), (strategy)>.
The process fragments of the enterprise are the following:
$PF_1$: <(customer inquiry), (order generation), (not forecasting strategy)>
$PF_2$: <(order generation), (goods issue), (not real time strategy)>
$PF_3$: <(goods issue), (goods delivery), (batch strategy)>
$PF_4$: <(goods delivery), (billing), (not data integrated strategy)>.

The process fragments of SAP are:
$PF_1$: <(customer inquiry), (order generation), (demand management strategy)>
$PF_2$: <(order generation), (goods issue), (material requirements planning strategy)>
$PF_3$: <(goods issue), (goods delivery), (inventory management strategy)>
$PF_4$: <goods delivery), (billing), (finance and control strategy)>.

Each from the above process fragments of SAP corresponds to a component (Table1).

| Process Fragments | SAP Components |
|---|---|
| PF1 | Demand management |
| PF2 | Master Production Scheduling<br>Material Requirements Planning |
| PF3 | Inventory Management |
| PF4 | Finance and Control |

Table 1: SAP Components.

From the comparison of the above process fragments we are coming into the following conclusions:

$PF_1$: The demand management module includes materials forecast so if demand management sub-module be implemented then the order management process of the enterprise will have the possibility to forecast requirements.

PF$_2$: With MPS and MRP components we can have interactive planning and date and time scheduling.

PF$_3$: With inventory management material stocks are recorded in the system according to both quantity and value with receipts and withdrawals, returns, transfers, and goods reservations being implemented. Inventory management is characterized by real time recording which enables checking and correction in the material flow [Keller and Teufel 1998].

PF$_4$: Finance and control offers data integration with sales, distribution and accounting so the enterprise can have accurate accounts.

Following the alignment process we know what to implement. We could have proceeded into a lower level through refinement. For example the MPS/MRP component could be:

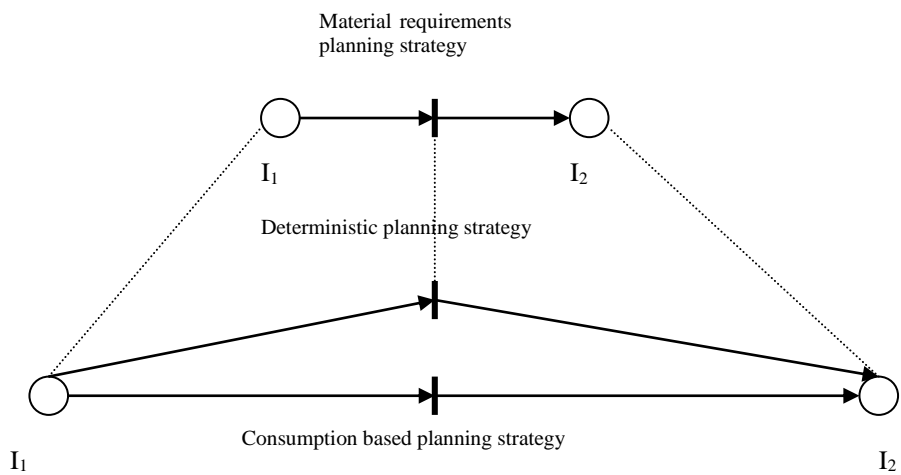

Figure14: Refinement of MRP sub-module

The planning strategy within MRP can be either deterministic or consumption based. In deterministic planning it is possible to decide whether and how materials are to be planned. Consumption based planning is based on a series of consumption values and is used together with forecast to plan future consumption.

The process fragment <(order generation), (goods issue), (deterministic planning strategy)> can be further refined into <(order generation), (goods issue), (gross planning strategy)>, <(order generation), (goods issue), (net planning strategy)>, <(order generation), (goods issue), (individual customer order planning strategy)>.

Accordingly the process fragment <order generation), (goods issue), (consumption based strategy)> could be refined into <(order generation), (goods issue), (reorder point strategy)>, <(order generation), (goods issue), (stochastic planning strategy)> etc.

The above could be proceeded further down. But because of the low level into which SAP functionality is described we remain into a higher level and thus the comparison is carried out with the models of fig.12 and fig.13.

## 4.1.3 Integrating Supply and Demand

Enterprise quest for integrating supply and demand was an old one and it began with its supply chain management initiative. The main goal is to implement "just-in-time" production scheduling, by dynamically updating manufacturing capacity and scheduling in response to continuously changing customer demands (both planned and unanticipated). The targeted processes are manufacturing resource planning (MRP-II), and more specifically, the Sales and Operations Planning (SOP) process within MRP-II.

The classic MRP-II integrates all business-related and logistics components. MRP-II takes into consideration all activities connected with the output of goods and services. It includes the following planning levels [Blain et al 1998]:
- Business Planning (operating results planning)
- Sales and Operation Planning
- Master Production Scheduling
- Material Requirements Planning
- Shop Floor Control

The starting point of the MRP-II planning chain is business planning, which usually turns production planning figures into money. The goal of sales and operation planning is to specify a production plan on the product group level (aggregate level). Using the aggregate production plan data, master production scheduling specifies the quantities to be produced of specific items in detail. Using data from the production plan, the MRP run produces either order proposals or direct vendor schedule releases. The shop floor control not only supports the creation and release, but also the monitoring of production orders.

Sales and Operations Planning is linking planning activities in upstream (manufacturing) and downstream (sales) operations. MRP-II process is illustrated in fig. 4.6 below.

In Petri-nets the sales and operations planning process of the enterprise (not the Sales and Operations component of SAP) will be:

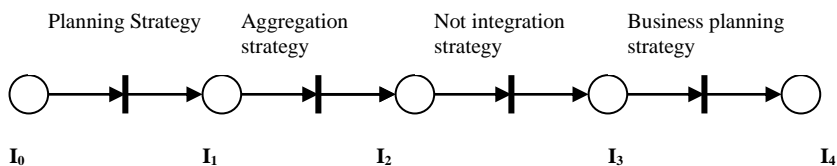

Figure 15: Enterprise Sales and Operations process

$I_0$: **order entry**
$I_1$: **requirements**
$I_2$: **requirements plan**
$I_3$: **proposed plan**
$I_4$: **final plan**

Prior to SAP, enterprise sales and operations planning was manual. In this approach, after the financial close of each month, sales planning and forecast data were aggregated from order entry and forecasting systems, validated, and manually keyed into master scheduling and production planning systems. In the same manner production and inventory data from the prior period were entered into order management systems. Two separate teams of demand analysts and supply planners were performing independent analysis and the results who could be different, were reconciled to determine the target production and target sales. Once an agreement was reached, in a business planning meeting senior executives were analysing the final production plan and demand schedule based on business assumptions, financial goals and other strategic parameters.

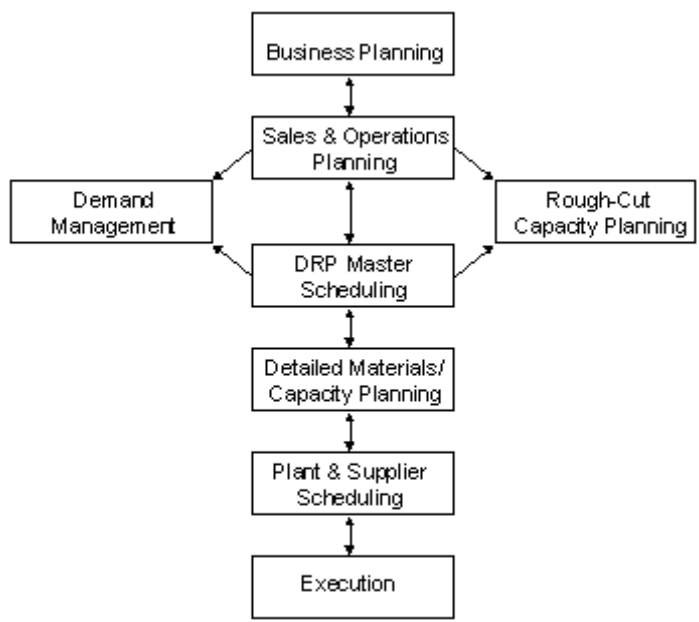

Figure16: Enterprise Manufacturing Planning Process [Battacharjee 2000].

While the manual SOP process was a major improvement compared to pre-SOP era, it was time consuming and error pruned in data re-entry and validation across sales production and financial systems. Furthermore the whole process took one month and it was not sensitive into changes and customer requests less than the period of one month. For example, once the plan for sales and production was issued, any additional request from any customer was considered after one month, when the new plan was issued.
The corresponding Sales and Operation process with SAP is:

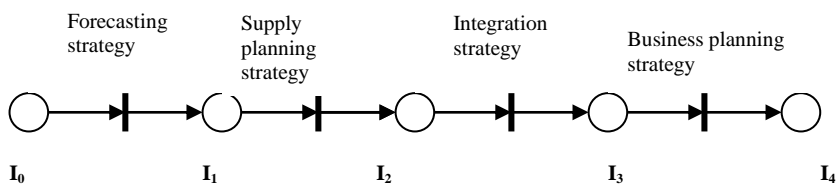

Figure 17: Sales and Operations Planning with SAP

$I_0$: order entry
$I_1$: requirements
$I_2$: requirements plan
$I_3$: proposed plan
$I_4$: final plan

The SOP (Sales and Operations Planning) component of Production Planning module combined with APO (Advance Purchase Organiser) could offer a reliable solution into enterprise problem with data analysis. The APO module includes the following components:
- Supply chain cockpit
- Forecasting
- Advance planning and scheduling
- Available to promise server (ATP).

In the same manner as in the previous section, we distinguish the process fragments of both models represented in fig. 16 and fig.17:

Enterprise:
$PF_1$<(order entry), (requirements), (planning strategy)>
$PF_2$<(requirements), (requirements plan), (aggregation strategy)>
$PF_3$<(requirements plan), (proposed plan), (not integration strategy)>
$PF_4$<(proposed plan), (final plan), (business planning strategy)>

SAP:
$PF_1$<(order entry), (requirements), (forecasting strategy)>
$PF_2$<(requirements), (requirements plan), (supply planning strategy)>
$PF_3$<(requirements plan), (proposed plan), (integration strategy)>
$PF_4$<(proposed plan), (final plan), (business planning strategy)>

The conclusions from the above comparison are:
$PF_1$: The planning strategy prior to SAP is mainly manual carried out by a number of meetings. In the case of SAP is fully automated carried out for example from the forecasting component of APO.
$PF_2$: Requirements prior to SAP were aggregated from order entry and forecasting systems and they were fed into production planning systems. In the case of SAP plans and schedules are derived live from status data, collected live from the system.
$PF_3$: Integration of supply and demand data was not real-time. Data were located in different databases. SAP supports real time integration of supply and demand data. For example SAP APO uses SAP LiveCache technology which processes data objects that are located in the memory.
$PF_4$: Business planning in the pre-SAP era carried out in a meeting from the senior management and it lasted one day. This lack business intelligence is fulfilled by the combination of SOP component together with the ATP component. Available To Promise (ATP) component can provide customers with accurate dates for complete or partial order fills.

The correspondence between process fragments of fig.17 and SAP components is:

| PROCESS FRAGMENTS | SAP COMPONENTS |
|---|---|
| PF1 | Forecasting (APO) |
| PF2 | Supply chain cockpit (APO) |
| PF3 | Advance planning and scheduling (APO) |
| PF4 | SOP & ATP |

Table 2

**Refinement**

Each from the above process fragments could be further refined and thus defining in more detail the reengineered process. For example:

The PF1 includes key activities such as product planning, sales planning and performance management for the prior period. More precisely this can be represented as:

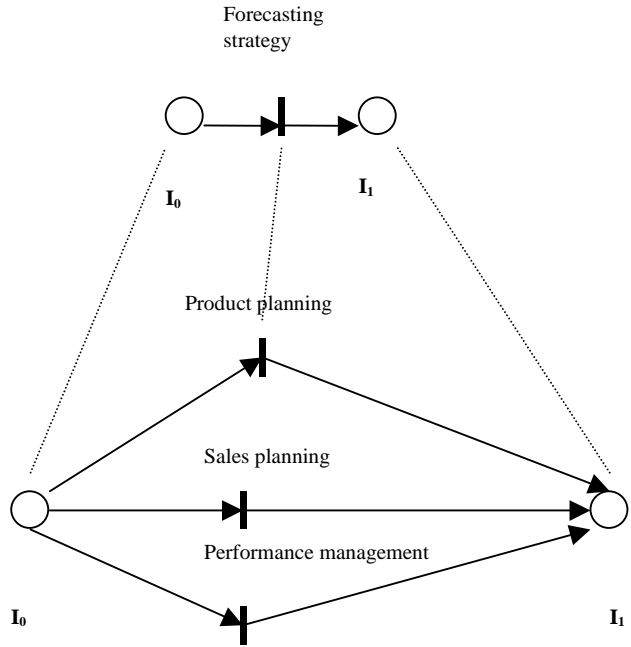

Figure 18: PF1 Refinement

**PF1 <(order entry), (requirements), (forecasting strategy)>**

⇩

**PF1.1 <(order entry), (requirements), (product planning)>, PF1.2<(order entry), (requirements), (sales planning)>, PF1.3<(order entry), (requirements), (performance management)>.**

The above notation explains what is described in the Petri-nets of fig.18.

Accordingly $PF_2$ could be refined as:

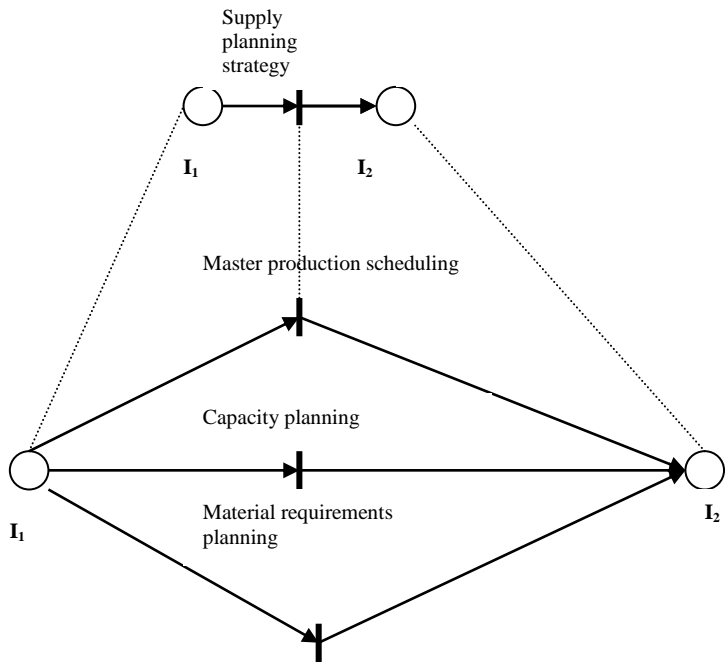

Figure 19: PF$_2$ Refinement

Figure 17 defines that: **PF2 <(requirements), (requirements plan), (supply planning strategy)>**

⇩

**PF2.1<(requirements), (requirements plan),(Master production scheduling)>, PF2.2<(requirements), (requirements plan), (capacity planning strategy), PF2.3<(requirements), (requirements plan), (material requirements planning)>.**

Process fragment PF3 <(requirements plan), (proposed plan), (integration strategy)> can be refined:

**PF3<(requirements plan), (proposed plan), (integration strategy)>**

⇩

**PF3.1<(requirements plan), (proposed plan), (consolidation strategy)> PF3.2<(requirements plan), (proposed plan), (feedback to demand and supply)>**

This can be visualised as follows:

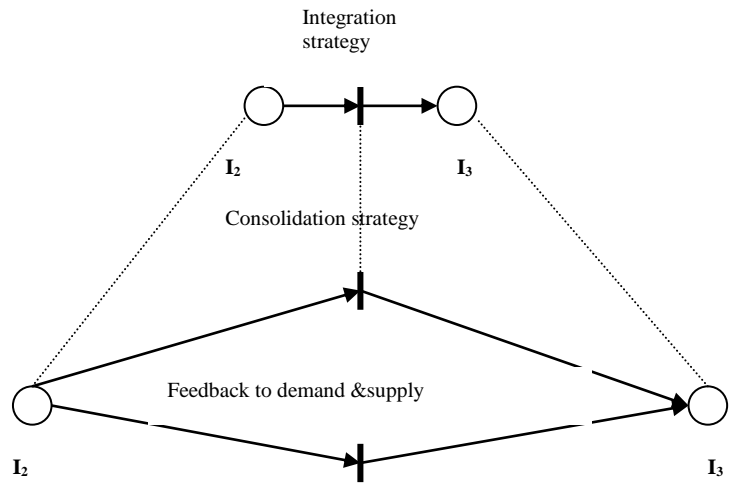

Figure 20: PF$_3$ Refinement

The process fragment (PF$_4$) <(proposed plan), (final plan), (business planning strategy)> can be refined in the same manner (fig. 21):

**PF4  <(proposed plan), (final plan), (business planning strategy)>**

⇩

**PF4.1<(proposed plan), (final plan), (performance review)>  PF4.2<(proposed plan), (final plan), (financial review)>**
**PF4.3<(proposed plan), (final plan), (approval/action items)>**

Figure 21: PF$_4$ refinement

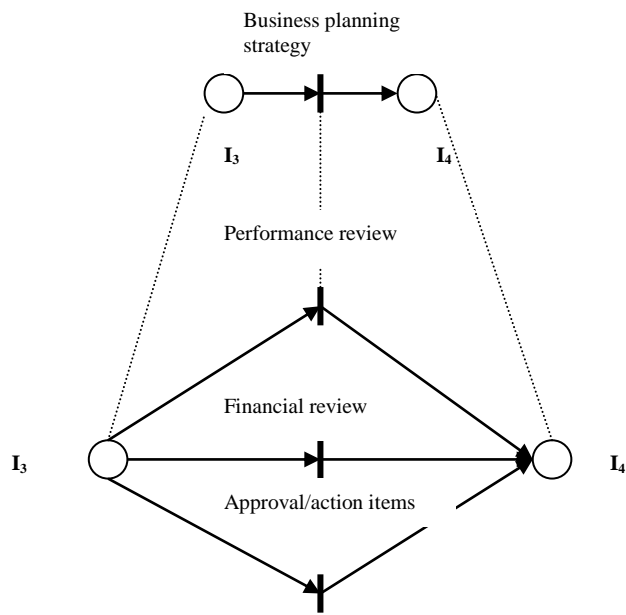

# 5. Conclusion

In this paper we have seen so far the ontology of the proposed methodology applied in to a Pharmaceutical Company.. The modeling of the current state of the enterprise and the proposed SAP solution is in Petri-nets, giving us the opportunity to represent the planning strategy from one state to another.

The levels of abstraction in each phase are three namely intentional, strategical, and operational. We think in terms of goals because organisations think in terms of their objectives [Rolland and Prakash 2000] and not in terms of their processes and functions. Supplementary to this we take into consideration the correspondent strategy in order to align the business processes of SAP to these of the enterprise.

SAP goals are supportive to enterprise goals, and are used to implement enterprise high level objectives such as increase of the profit, which is an objective of every commercial organisation, as well as, improvement of the customer satisfaction, improvement of the quality of service and as a result to this, consistent growth and increase of the market shares. Perhaps every commercial organisation has to choose within a variety of computer systems solutions but when it chooses ERP and more specifically SAP, it must have the possibility of gaining the most from this choice. .

ERP systems are often large systems, of many interacting components. Each component interacts with other components within the system. Thus despite the diversity of the systems we want to model, several common points stand out. These points are modularity and concurrency. Modularity because systems consisting of separate interacting components. Each component may itself be a system and its behaviour can be described independently from other components, apart from well-defined interactions. Concurrency because activities of one component of a system may occur simultaneously with other activities of other components.
The conclusions are summarized in the following: Firstly the way of approach in a real project. SAP is not being developed as a unity, but we have a targeted process. For example, in the first case study, we don't develop the PP module with all the components. An adjustment in a targeted process is being done and all the necessary components are implemented. In the case of the demand side implementation only one component from the PP module was implemented. So the intention is not to develop SAP itself, but to adjust all its necessary components to facilitate the implementation of a targeted process.

The above conclusions can be represented in a metamodel, the SAP implementation method metamodel.
Central to this metamodel is the SAP Implementation method entity. This is depended upon the targeted process which every SAP project realizes and the enterprise type. By the type we mean the nature of the enterprise and therefore the appropriate SAP project type. For example the enterprise can be located in a single plant or distributed within differently located plants possibly even to alternate continents.

The enterprise is ruled by the stakeholders who determine enterprise goals. On the other hand supportive to enterprise goals are SAP goals. These are realized by the SAP functions. Finally SAP structure is depicted by the presence of the entities module and functions. Secondly the applicability of the framework is satisfactory. We have a very good representation of the current situation of the enterprise as well as the future's solution with SAP. The comparison or better the alignment (validation phase) is carried out in a high strategy level but we can proceed further down in a lower level through refinement. Finally a correspondence of the process fragments and components of SAP is carried out as a result of the proposed solution with SAP and the implementation aspects of the project. In the second case study the correspondence is between fragments and SAP modules.

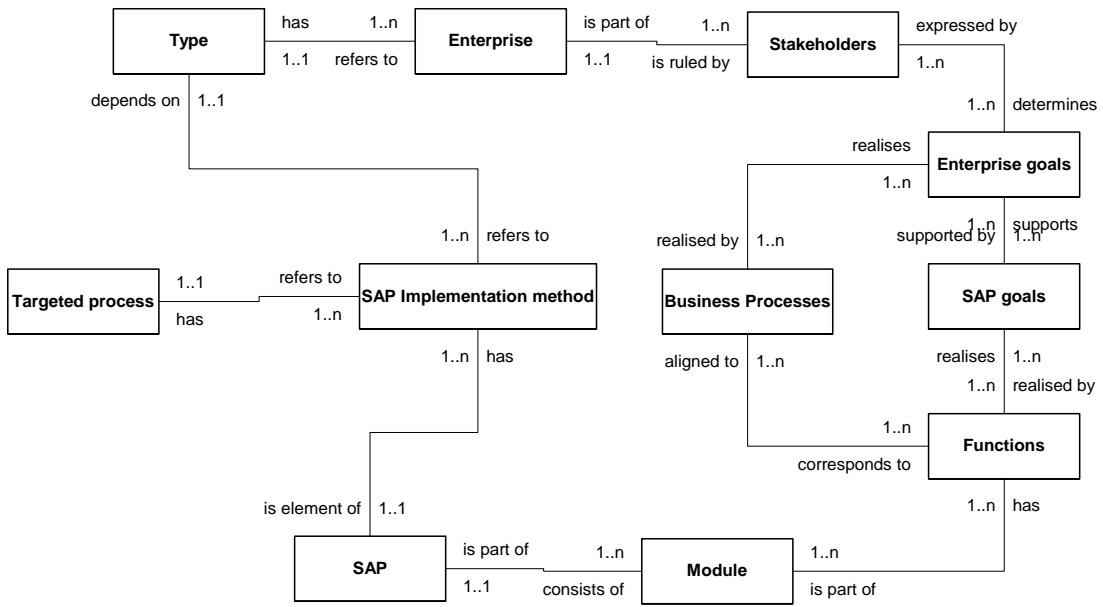

Figure 22: SAP Implementation method metamodel

# 6. References

**Aamodt, A., Plaza, E.** Case-based reasoning: Foundations issues, methodological variations and system approaches. *AI-Communications*, 7(1), 1994, pp 39-59.

**Battacherjee, J.** SAP R/3 Implementation at Geneva Inc. JCAIS, Communications of the Association of Information Systems. Vol. 4 (3), 2000.

**Blain et al.** Using SAP R/3. Que, 1998.

**Buck-Emden, R.** The SAP R/3 System, An Introduction to ERP and Business Software Technology. Addison-Wesley, 2000.

**Hiquet, B**. *SAP R/3 Implementation Guide*. Macmillan Technical Publishing, 1998.

**Keller, G., Teufel, T**., *SAP R/3 Process Oriented Implementation.* Addison-Wesley, 1998.

**Kolezakis, M., Loucopoulos, P.** *Alignment of the SAP requirements to Enterprise requirements?. EMMSAD '03, CAiSE 2003, Velden Austria, 16-17 June.*

**Loucopoulos, P., Karakostas, V**. *Systems Requirements Engineering*. McGraw-Hill, 1995.

**Loucopoulos, P., Kavakli, V.** Enterprise Knowledge Management and Conceptual Modelling, ER 97, 1997

**Peterson, J.L** Petri Net *Theory and the modelling of Systems*. Prentice-Hall, 1981.

**Petri, C**. *Communication with Automata*. New York:Griffiss Air Force Base. Tech. Rep. RADC-Tr-65-377, vol.1, Suppl. 1.

**Ramesh, B., Jarke, M.,** Towards Reference models for Requirements Traceability. *TSE* 27(1) 58-93, 2001.

**Rolland, C., Prakash**, N. Bridging The Gap Between Organisational Needs And ERP Functionality. *Requirements Engineering* 5(3), 2000, pp180-193.

**Scheer, A-W**. *Business process engineering: reference models for industrial enterprises*. Berlin et al., 1994.

**Weihrauch, K.** SAPinfo-Continous Business Engineering, Waldorf, 1996.

**Welti, N.** Successful SAP R/3 Implementation, Practical Management for ERP Projects. Addison-Wesley, 1999.